\documentclass[12pt]{iopart}
\usepackage{graphicx}
\usepackage{cite}
\usepackage{eqnarray}
\usepackage[colorlinks,
linkcolor=blue,
anchorcolor=blue,
citecolor=blue,
urlcolor=blue
]{hyperref}
\usepackage{todonotes}

\begin{document}
\title[]{Quantitative 3D non-linear simulations of shattered pellet injection in ASDEX Upgrade using JOREK}

\author{W. Tang$^1$, M. Hoelzl$^1$, P. Heinrich$^1$, D. Hu$^2$, F. J. Artola$^3$, \\P. de Marn\'{e}$^1$, M. Dibon$^3$, M. Dunne$^1$, O. Ficker$^4$, \\P. Halldestam$^1$, S. Jachmich$^3$, M. Lehnen$^{3}$\footnote[1]{Deceased}, E. Nardon$^5$, \\G. Papp$^1$, A. Patel$^1$, U. Sheikh$^6$, the ASDEX Upgrade Team$^a$, the EUROfusion Tokamak Exploitation Team$^b$ and the JOREK Team$^c$}            

\address{$^1$Max Planck Institute for Plasma Physics, Garching, Germany\\
         $^2$Beihang University, Beijing, China\\
         $^3$ITER Organization, Saint-Paul-Lez-Durance, France\\
         $^4$Institute of Plasma Physics of the CAS, Prague, Czech Republic\\
         $^5$CEA/IRFM, Saint-Paul-Lez-Durance, France\\
         $^6$\'{E}cole Polytechnique F\'{e}d\'{e}rale de Lausanne (EPFL), Swiss Plasma Center (SPC), Lausanne, Switzerland\\
         $^a$See the author list of H. Zohm et al. 2024 \textit{Nucl. Fusion} \textbf{64} 112001\\
         $^b$See the author list of E. Joffrin et al. 2024 \textit{Nucl. Fusion} \textbf{64} 112019\\
         $^c$See the author list of M. Hoelzl et al. 2024 \textit{Nucl. Fusion} \textbf{64} 112016
         }
\ead{weikang.tang@ipp.mpg.de}
\vspace{10pt}

\begin{abstract}
    Shattered pellet injection (SPI) as primary mitigation method for major disruptions in ITER has a large parameter space available for optimization including the total amount of injected material, the size of the individual pellet fragments, the material composition, and the timing of multiple injections. This flexibility needs to be exploited to simultaneously minimize thermal heat loads, electromagnetic vessel forces, and formation of relativistic electrons and their impacts on plasma facing components. In this article, we apply 3D non-linear magnetohydrodynamic modelling to SPI experiments in the ASDEX Upgrade tokamak, going beyond our previous work [Tang et al \textit{Nucl. Fusion} \textbf{65} 116003 (2025)] by resolving some discrepancies between simulations and experiment and thus opening the path to quantitative model validation and experiment interpretation. The key element that enables the transition from merely qualitative comparisons to quantitatively reliable predictions of the thermal quench duration and the radiation fraction is the incorporation of a simplified treatment of parallel heat-flux limiting. The work increases the confidence of matching the key processes of disruption mitigation with this high fidelity modelling in view of predictive studies for ITER.
\end{abstract}

\vspace{2pc}


\maketitle


\section{Introduction}\label{sec1}

    Major disruptions in large tokamak devices pose serious challenges to the lifetime of plasma facing components and machine integrity~\cite{Boozer2012,Bandyopadhyay_2025}. Shattered pellet injection (SPI) is among the most promising mitigation strategies~\cite{SPIORNL,Gebhart_2021}, rapidly delivering massive amounts of material into the plasma to radiate energy uniformly and suppress runaway electron formation~\cite{Breizman_2019}. To effectively mitigate disruption damage, ITER has chosen SPI as the baseline of its disruption mitigation system (DMS)~\cite{Lehnen_2015,DMSIAEA}. Accurate predictive modeling of SPI is therefore essential to optimize injection parameters and evaluate mitigation performance for the ITER DMS.

    In collaboration with ITER, a flexible SPI system with three independent injection lines was installed on the ASDEX Upgrade (AUG) tokamak~\cite{Dibon,HEINRICH2024114576,Heinrich_2025_PhD}. During the 2022 AUG campaign, a dedicated series of experiments was conducted with the goal of identifying the optimal fragment configuration for effective disruption mitigation~\cite{paul2025nf_frad,Heinrich_2025_PhD}. To support this effort, extensive simulation activities using various numerical codes were undertaken, providing crucial input for interpreting and validating the experimental observations~\cite{Schwarz_2023, peter_augspi,ansh_augspi,tang2025augspi,Oskar2025PPCF}. 
    
    In our previous work~\cite{tang2025augspi}, 3D non-linear magnetohydrodynamic (MHD) simulations of SPI into AUG H-mode plasmas were successfully performed with various injection scenarios using the JOREK code~\cite{Huysmans2007,jorek2024}. However, the simulated pre-thermal quench (pre-TQ) phase was found to be much shorter than that observed in AUG experiments~\cite{Heinrich_2025_PhD, Jachmich_2023_EPS, Jachmich_2024_EPS}. In our earlier simulations, parallel thermal diffusivity was modeled using the Spitzer–Härm formula~\cite{SHdiff}, which is valid under the assumption of a collision-dominated plasma, where the electron mean free path $\lambda_e$ is much smaller than the electron temperature scale length $L_T=T_e/\nabla T_e$. During an SPI induced TQ, strong local cooling created by the cryogenic pellet and ablated impurity radiation can lead to locally steep $\nabla T_e$ along the magnetic field lines. Thus, the Spitzer–Härm formula might significantly overestimate the parallel heat flux, beyond the free-streaming (saturated) limit $q_\parallel~\sim~\alpha n_eT_ev_{th}$, where $q_\parallel$ is the parallel heat flux, $v_{th}$ is the electron thermal speed and $\alpha\approx0.1~-~0.3$~\cite{Tokar_2007,Bale_2013}. Kinetic simulations indicate that during a plasma thermal quench, energy loss proceeds via a self-regulated kinetic cooling flow whose magnitude is limited by ambipolar electric fields and return currents, naturally producing saturated parallel heat fluxes well below the free-streaming limit~\cite{Zhang_2023}. Experiments have shown that electron heat conduction is reduced to values far below classical predictions and heat flux limit needs to be applied~\cite{Malone75}. Numerical studies also suggest that the thermal diffusivity is lower than the Spitzer–Härm prediction by one order of magnitude in TEXTOR experiments~\cite{Matthias2009}. 
  
    In this paper, we revisit the previous SPI simulations and investigate the effect of parallel thermal diffusivity reducing it by a constant factor as a simplified way of including the heat flux limit. We find that a reduction by exactly a factor of ten yields pre-TQ durations in quantitative agreement with experiments, without degrading agreement in other aspects of the disruption. The rest of the paper is structured as follows: the simulation setup is briefly introduced in section~\ref{sec2}. In section~\ref{sec3}, we first investigate the effect of parallel thermal diffusivity $\chi_{\|}$ based on one of our previous cases from Ref.~\cite{tang2025augspi}. Then the influence of neon fraction and fragment size is re-assessed based on the results with reduced $\chi_{\|}$ including direct experiment comparisons.

\section{Simulation setup}\label{sec2}

    The two-temperature reduced MHD model used in this study is identical to that described in section 2.1 of Ref.~\cite{tang2025augspi}. The equilibrium is reconstructed from AUG H-mode discharge \#40355, with a toroidal field $B_t$~=~1.8~T and a plasma current $I_p$~=~0.8~MA. Unless otherwise stated, the simulation parameters are the same as those described in section 3.1 of Ref.~\cite{tang2025augspi}, except that the Spitzer–Härm parallel thermal diffusivity is reduced by multiplying it with a factor of 0.1 to account for flux-limiting effects, i.e. 
    
\begin{equation}
\chi_\|=0.1\cdot\chi_{\|, S H}=0.1\cdot3.6\times 10^{29} \frac{T_e[\mathrm{keV}]^{5 / 2}}{n_{e}\left[m^{-3}\right]} m^2 / \mathrm{s}.
\end{equation}

\begin{figure}[htb!]
\centering
\includegraphics[width=1.\textwidth]{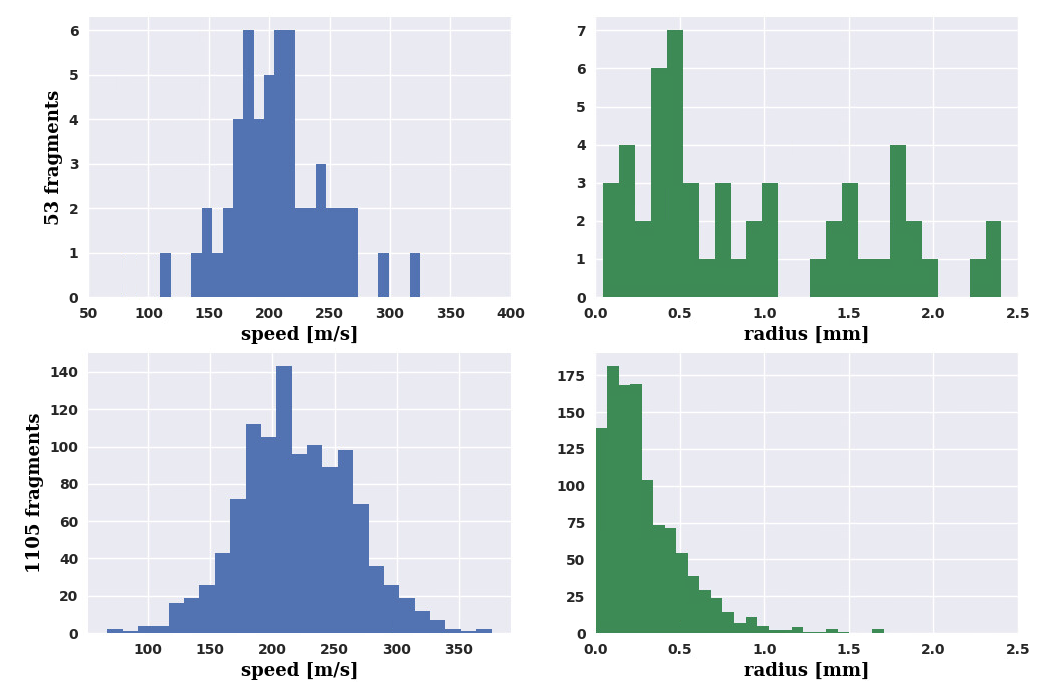}
\caption{Speed and radius distributions of the fragments. The first row illustrates the distributions for the 53 fragment case (same as the LF\_HV\_Ne10 case in~\cite{tang2025augspi}), used to compare the effects of different $\chi_{\|}$ values and neon fractions. The second row illustrates the distribution of the 1105 fragment case, which has a similar speed distribution but different size distribution  (both sampled based on Parks' fragmentation model~\cite{parks}) compared to the 53 fragment case, and is used to study the effect of different fragment size.}
\label{fig_dist}
\end{figure} 
    
    The SPI is launched at the very beginning of our simulations. Once a fragment enters the plasma domain, the ablation rate for a mixed neon and deuterium fragment is calculated based on the neutral gas shielding (NGS) model~\cite{parksabl}. The impurity atomic processes are modeled by loading marker particles with their collisional-radiative rate obtained based on the OpenADAS database~\cite{Hu2021b}. No background impurities or initial magnetic perturbations are included, so that the simulations focus solely on the effects of the injected material. Earlier work suggests that this is well justified unless pure deuterium injection is studied~\cite{Hu_2023}. 
        
\section{Numerical results}\label{sec3}

    In this section, we first investigate the effect of parallel thermal diffusivity $\chi_{\|}$ based on the case LF\_HV\_Ne10 of Ref.~\cite{tang2025augspi}, corresponding to a 10\% neon pellet with a pre-shatter speed of 222~m/s shattered by a 25$^\circ$ rectangular cross-section shatter head. Based on the results with reduced $\chi_{\|}$, we then re-assess the influence of neon fraction and fragment size. The speed and radius distributions for the cases simulated in this work are shown in figure~\ref{fig_dist}.

\begin{figure}[htb!]
\centering
\includegraphics[width=.8\textwidth]{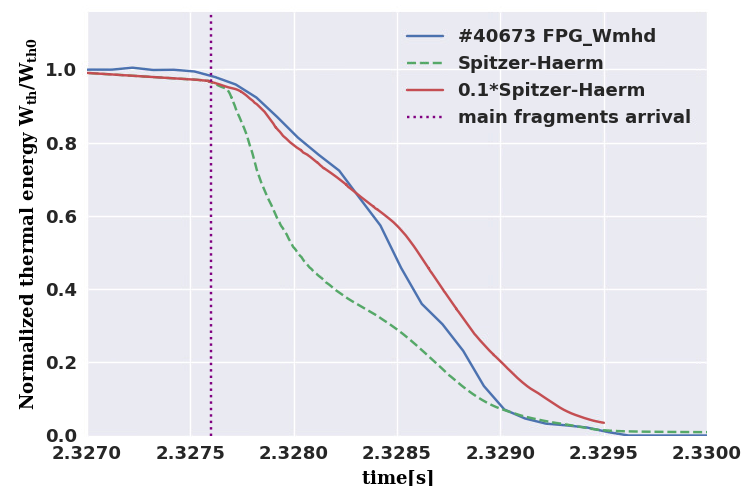}
\caption{Evolution of the thermal energy for the baseline simulation with $\chi_{\|,SH}$, the new case with reduced parallel thermal diffusivity $0.1\cdot\chi_{\|,SH}$, in comparison with the experiment discharge \#40673. The thermal energy signal in experiment FPG\_Wmhd is reconstructed through function parametrization~\cite{Braams_1986_FP,McCarthy_1992_PhD}. In this particular experimental discharge, the pellet speed is 221 m/s, the shatter angle is 25 degrees, and the neon fraction is 10\%. The traces from the simulation are time-shifted so that the fragment arrival time coincides with that of the experiment, which is annotated by the purple dotted line.}
\label{fig_zkpar_th}
\end{figure} 

\subsection{Effect of parallel thermal diffusivity}

    To begin with, figure~\ref{fig_zkpar_th} shows the time evolution of the thermal energy for the baseline simulation with $\chi_{\|,SH}$, the new case with reduced parallel thermal diffusivity $0.1\cdot\chi_{\|,SH}$, and the corresponding experimental measurement. The baseline simulation features a too steep and strong initial loss of thermal energy after the fragment arrival when compared to the experiment. By contrast, the simulation with reduced $\chi_{\|}$ exhibits much closer quantitative agreement with the experimental data, indicating that the nominal Spitzer–Härm value overestimates parallel heat transport in this regime.

\begin{figure}[htb!]
\centering
\includegraphics[width=1.\textwidth]{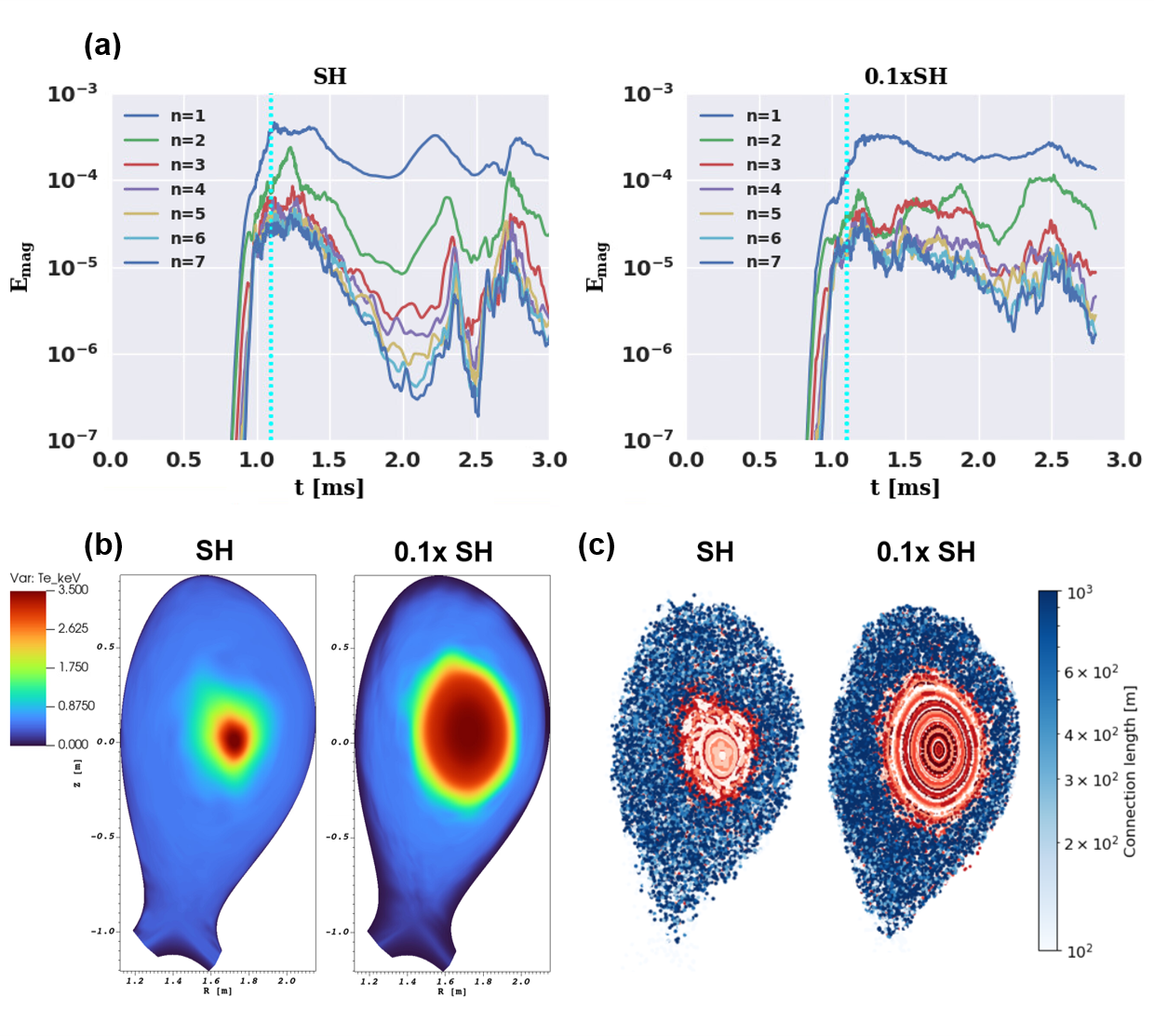}
\caption{(a) Non-linear evolution of the MHD magnetic energy spectra for the cases with $\chi_{\|,SH}$ and $0.1\cdot\chi_{\|,SH}$. (b) Comparison of electron temperature $T_e$ and (c) Poincar\'e plot at the injection plane at simulation time 1.1~ms, corresponding to the time marked by the cyan dotted line in (a). In (c), red markers represent confined field lines, while blue markers indicate those lost to the boundary.}
\label{fig_tepoin}
\end{figure} 

    Figure~\ref{fig_tepoin}~(a) compares the MHD spectra for the above mentioned cases with different $\chi_{\|}$. We show that the $n~=~1$ and $n~=~2$ harmonics (note the logarithmic scale) are much stronger in the case with $\chi_{\|,SH}$ during the initial phase after pellet injection, although the $n=1$ component remains the dominant harmonic in both cases, where $n$ is the toroidal mode number. This indicates a significantly stronger MHD response when a larger $\chi_{\|}$ is used. 

\begin{figure}[htb!]
\centering
\includegraphics[width=1.\textwidth]{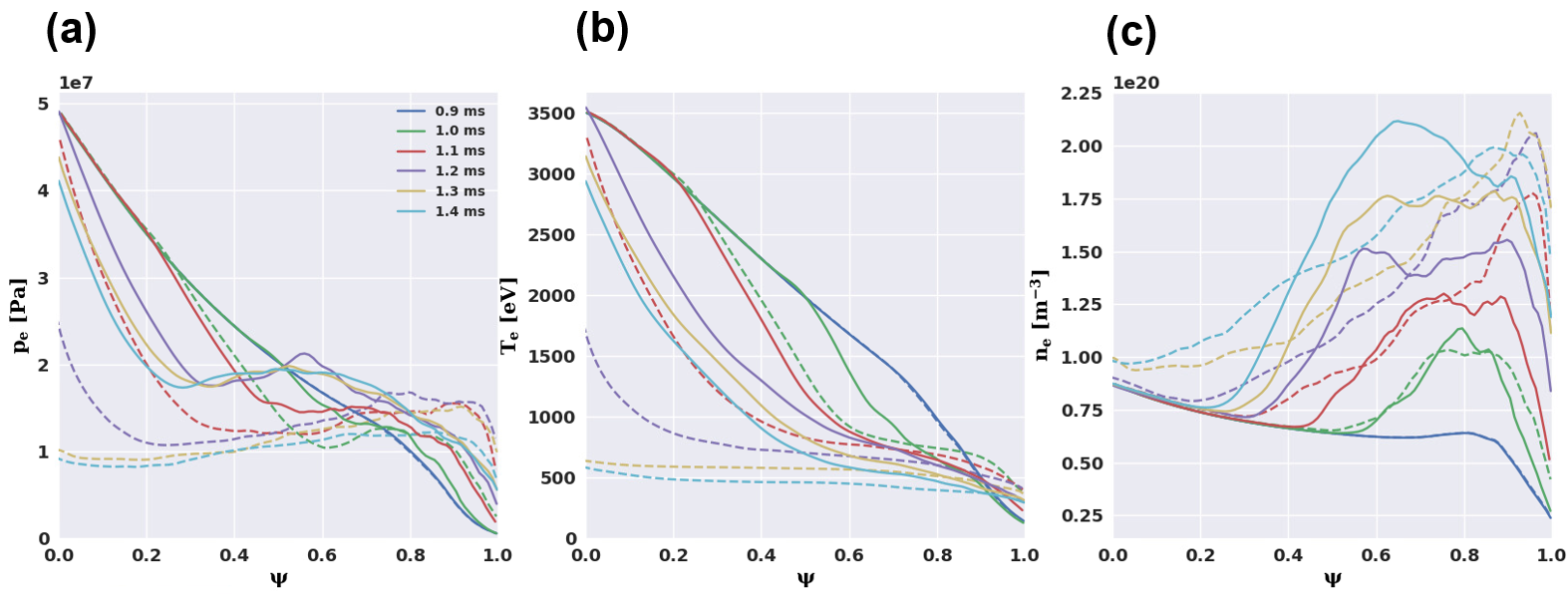}
\caption{Profiles of (a) electron pressure $p_e$, (b) electron temperature $T_e$ and (c) electron density $n_e$ at different simulation times. The dashed lines show the evolution for the case with $\chi_{\|,SH}$ and the solid lines display the case with $0.1\cdot\chi_{\|,SH}$.}
\label{fig_drift}
\end{figure} 

    There are few mechanisms through which an overestimated parallel thermal diffusivity can artificially enhance the thermal energy loss in the simulations. First, the overestimated parallel heat flux results in too fast re-heating of the adiabatically forming neutral cloud, which exaggerates the plasmoid-drift-induced losses~\cite{Hu_2024,plasmoid_kong,Pegourie_2007}. Second, excessive parallel heat transport strengthens cooling along magnetic field lines, leading to a rapid increase in plasma resistivity. The elevated resistivity, in turn, facilitates the growth of resistive MHD instabilities~\cite{Nardon_2017}, thereby amplifying the overall MHD response.  Finally, in regions where the magnetic topology becomes stochastic, the overestimated thermal diffusivity causes unrealistically high thermal energy losses, particularly during the initial cooling stage. This excessive energy depletion results in a shorter and more abrupt pre-TQ phase than that observed in the experiments. Figures~\ref{fig_tepoin}~(b) and (c) further compare the electron temperature $T_e$ and the Poincar\'e plots at the injection plane, taken approximately 0.2~ms after the fragments arrival ($t~=~1.1$~ms). With the reduced $\chi_{\|}$, the core flux surfaces remain more intact, indicating milder MHD destabilization. The connection length is also longer, leading to less conductive heat loss at the plasma edge.

\begin{figure}[htb!]
\centering
\includegraphics[width=1.\textwidth]{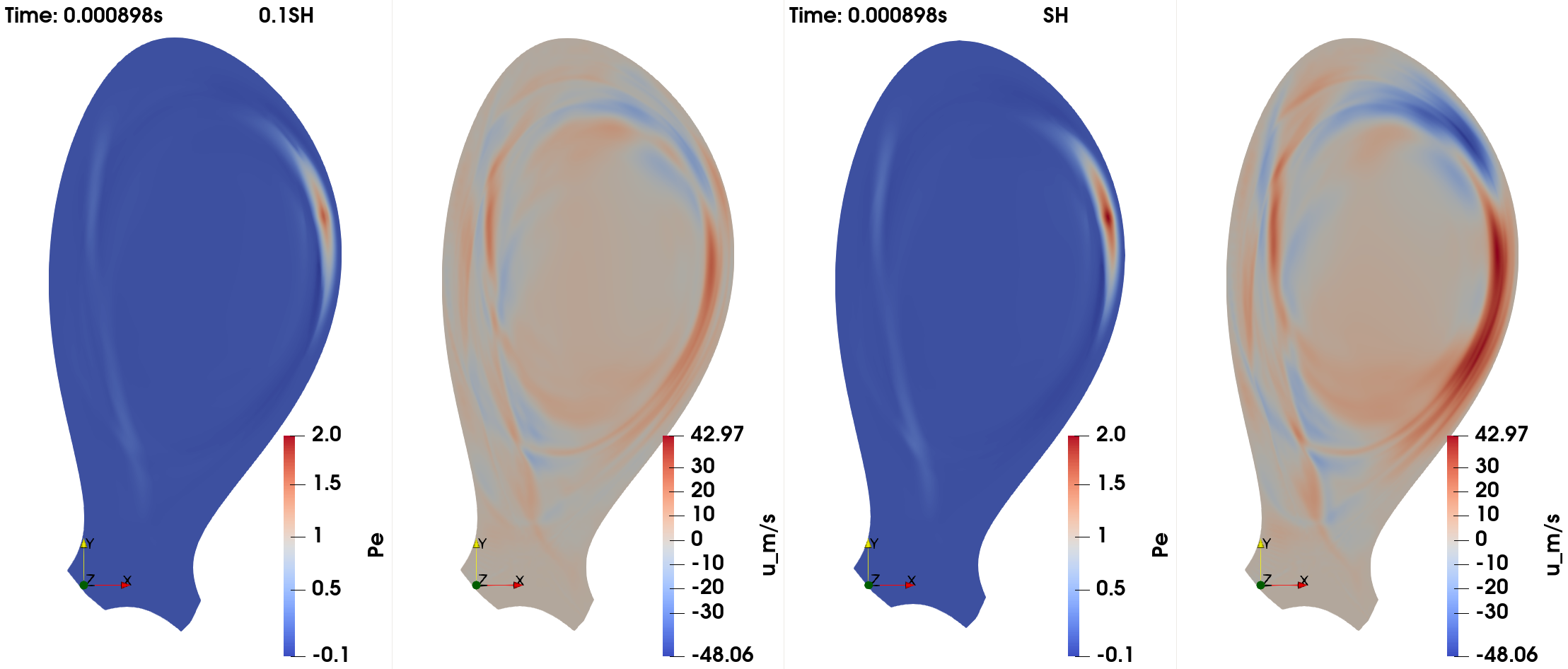}
\caption{The perturbed electron pressure $\delta p_e=p_e-p_{e0}$ and non-axisymmetric part of stream function $u$ (electric potential) at simulation time 0.9~ms, for the case with $0.1\cdot\chi_{\|,SH}$ (left) and the case with $\chi_{\|,SH}$ (right).}
\label{fig_drift2}
\end{figure} 

\begin{figure}[htb!]
\centering
\includegraphics[width=.9\textwidth]{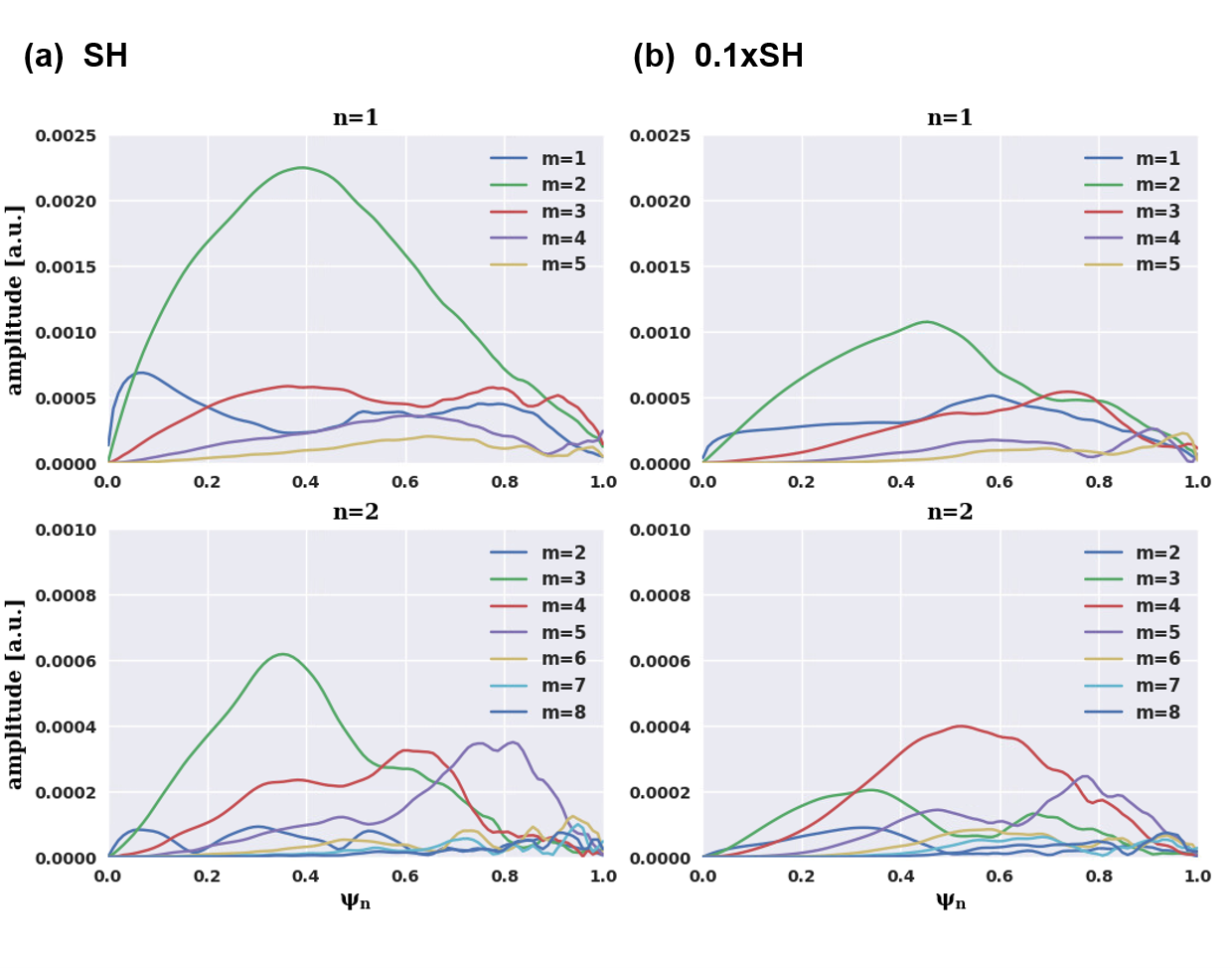}
\caption{Radial structure of the perturbed poloidal magnetic flux $\delta\psi$ decomposed into different mode numbers for (a) $\chi_{\|,SH}$ and (b) $0.1\cdot\chi_{\|,SH}$ at simulation time 1.1~ms. The first row shows the toroidal mode number $n=1$ with poloidal mode numbers $m=1$~–~$5$, while the second row shows $n=2$ with $m=2$~–~$8$.}
\label{fig_fft}
\end{figure} 

    Figure~\ref{fig_drift} shows the flux-averaged profiles of electron pressure $p_e$, electron temperature $T_e$ and electron density $n_e$ from 0.9 to 1.4~ms, for the cases using $\chi_{\|,SH}$ (dashed) and $0.1\cdot\chi_{\|,SH}$ (solid). Comparing the $T_e$ profiles, we show that the the cold front is not fully established during this time window and the temperature difference is is not sufficiently large for the resistivity mechanism discussed above to play a dominant role. Therefore, the exaggerated MHD response observed here is primarily attributed to the plasmoid drift mechanism discussed previously. In figure~\ref{fig_drift2} the change of local pressure $\delta p_e$ and stream function $u$ (electric potential) at the injection plane is compared, for the cases with $0.1\cdot\chi_{\|,SH}$ and $\chi_{\|,SH}$. The comparison shows that, with the larger $\chi_{\|}$, the plasmoid is heated more rapidly through parallel heat flux, leading to a higher local pressure. This elevated pressure enhances charge separation due to the $\nabla B$ drift, as reflected in the plot of $u$, and ultimately results in a stronger major-radial plasmoid drift. As the drift develops, plasmoids drag magnetic field lines with them, generating perturbed magnetic fields $\delta \mathbf{B}$ and thereby destabilizing MHD instabilities.

\begin{figure}[htb!]
\centering
\includegraphics[width=1.\textwidth]{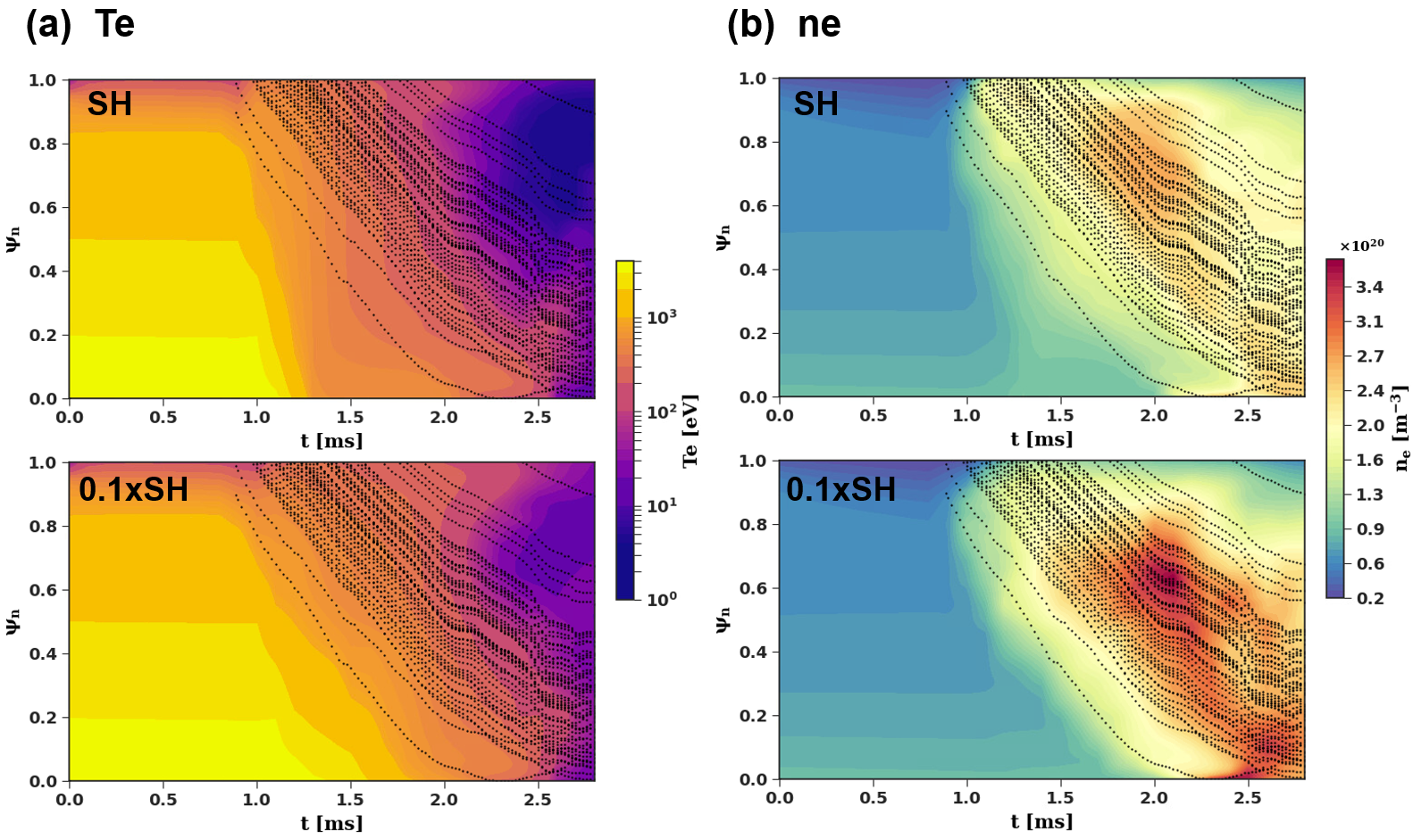}
\caption{Flux-averaged (a) electron temperature $T_e$ and (b) electron density $n_e$ profile evolution for the cases with $\chi_{\|,SH}$ and $0.1\cdot\chi_{\|,SH}$. The dotted traces indicate the trajectories of all the fragments (including the already fully ablated fragments), evolved according to their initial velocities.}
\label{fig_zkpar_teprof}
\end{figure} 

    To examine the MHD mode structures in more detail, figure~\ref{fig_fft} displays the radial profiles of the perturbed poloidal magnetic flux $\delta\psi$ for different mode numbers $m/n$ at $t~=~1.1$~ms. Comparing the $n~=~1$ and $n~=~2$ components, it is evident that the dominant mode is $n~=~1$, with the main component being the $m/n~=~2/1$ mode at that time. The mode amplitude in the case with $\chi_{\|,SH}$ is significantly larger than that in the $0.1\cdot\chi_{\|,SH}$ case, and the mode structure exhibits the typical characteristics of a resistive tearing mode, indicating that the growth rate of 2/1 tearing modes is greatly amplified through the aforementioned mechanism when fragments arrive at the $q~=~2$ surface. Additionally, when examining the $n~=~2$ component, a more pronounced $m/n~=~3/2$ mode appears in the $\chi_{\|,SH}$ case. This secondary mode arises mainly from non-linear mode coupling. As shown in figure~\ref{fig_tepoin}~(a), this $m/n~=~3/2$ mode can further contribute to the rapid thermal collapse at a later stage, when it grows to an amplitude comparable to the $n~=~1$ mode at about 1.2~ms.

    Figure~\ref{fig_zkpar_teprof} illustrates the time evolution of the flux-averaged electron temperature $T_e$ and electron density $n_e$ profiles for the cases with $\chi_{\|,SH}$ and $0.1\cdot\chi_{\|,SH}$. By comparing the $T_e$ evolution, it is evident that with $\chi_{\|,SH}$, the cooling front propagates significantly ahead of the fragment trajectory. The core $T_e$ collapses rapidly while the fragments are still located near the plasma edge, indicating an overestimated MHD response in this case. In contrast, for the simulation with $0.1\cdot\chi_{\|,SH}$, the cooling front remains more closely aligned with the fragment trajectory, suggesting what we believe to be a more realistic representation of the thermal diffusivity. Regarding the $n_e$ evolution, the ablation front is well aligned with the fragment position in both cases. For the reduced $\chi_{\|}$ case, a considerably higher ablation fraction is reached due to the higher local temperature at the fragment locations maintained by the slower thermal collapse.

\begin{figure}[htb!]
\centering
\includegraphics[width=.8\textwidth]{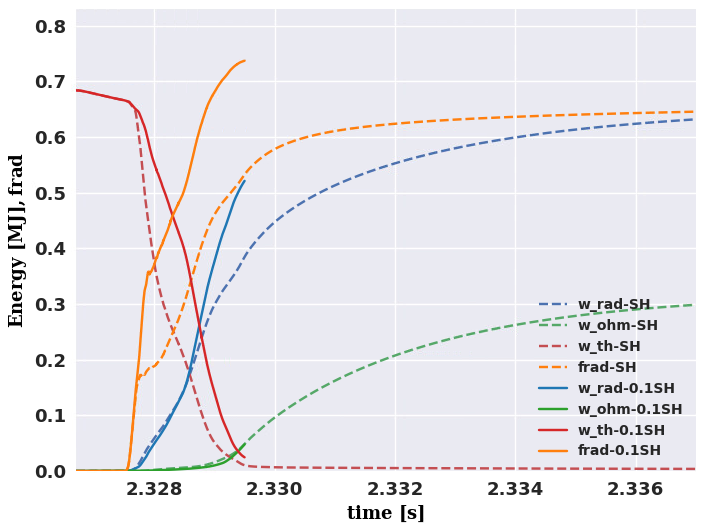}
\caption{Time evolution of total radiated energy, ohmic heating energy, total thermal energy and the radiation fraction for the case with $\chi_{\|,SH}$ (dashed) and $0.1\cdot\chi_{\|,SH}$ (solid). The blue traces represent the radiated energy $W_{rad}$, the green traces show the ohmic heating energy $W_{ohm}$, the red traces display the thermal energy $W_{th}$, and orange traces the radiation fraction. The radiation fraction is calculated by $frad(t)=W_{rad}(t)/[W_{th}(0)-W_{th}(t)+W_{ohm}(t)]$.}
\label{fig_zkpar_frad}
\end{figure} 

\begin{figure}[htb!]
\centering
\includegraphics[width=.9\textwidth]{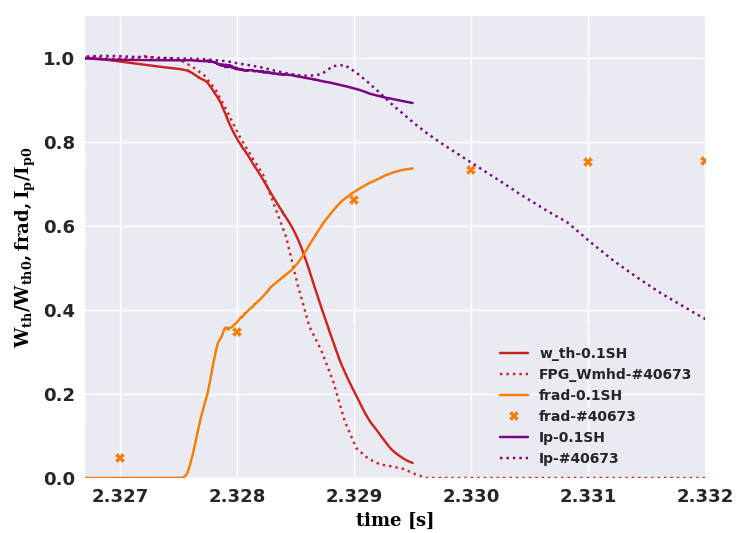}
\caption{Time evolution of thermal energy, plasma current and radiation fraction for injections of 10\% neon pellets with $0.1\cdot\chi_{\|,SH}$, in comparison with the corresponding experiment discharge \#40673 (10\% neon, pellet speed 221~m/s, $25^\circ$ shatter angle). The thermal energy and plasma current are normalized to the values at t=0 for a clear comparison. The `X' marker shows the radiation fraction from experimental data~\cite{paul2025nf_frad}.}
\label{fig_exp_40673}
\end{figure} 

    Figure~\ref{fig_zkpar_frad} shows the time evolution of the total radiated energy, ohmic heating energy, total thermal energy, and radiation fraction for the cases with $\chi_{\parallel,SH}$ and $0.1\cdot\chi_{\parallel,SH}$. We show that in the case with $\chi_{\parallel,SH}$, the radiation fraction during the initial cooling stage is very limited due to the overestimated heat losses. This leads to a relatively lower final radiation fraction compared to the experimental results, as described in Ref.~\cite{tang2025augspi}. Although the total radiated energies in the early phase are comparable between the two cases, the initial thermal energy drop in the case with $\chi_{\parallel,SH}$ is significantly steeper, indicating that majority of the thermal energy loss in the $\chi_{\parallel,SH}$ case occurs through conduction and convection rather than radiation in the early phase. By contrast, when $0.1\cdot\chi_{\parallel,SH}$ is applied, the simulation results show a much higher radiation fraction. 

    The simulation result with $0.1\cdot\chi_{\parallel,SH}$ is compared with its experimental counterpart in figure~\ref{fig_exp_40673}, in terms of thermal energy, plasma current and radiation fraction. Numerical results show a quantitative agreement with experimental data, especially in thermal energy evolution and radiation fraction. While the pronounced current spike observed in the experiment is not reproduced in the simulation, the early phase of the plasma current decay is captured reasonably well.  

\subsection{Revisited the influence of neon content and fragment size}

    The above results suggest that the simulation fidelity can be significantly enhanced by applying the reduced $\chi_{\parallel}$. In this subsection, we revisit the effects of neon fraction and fragment size based on the reduced $\chi_{\parallel}$.

\begin{figure}[htb!]
\centering
\includegraphics[width=.8\textwidth]{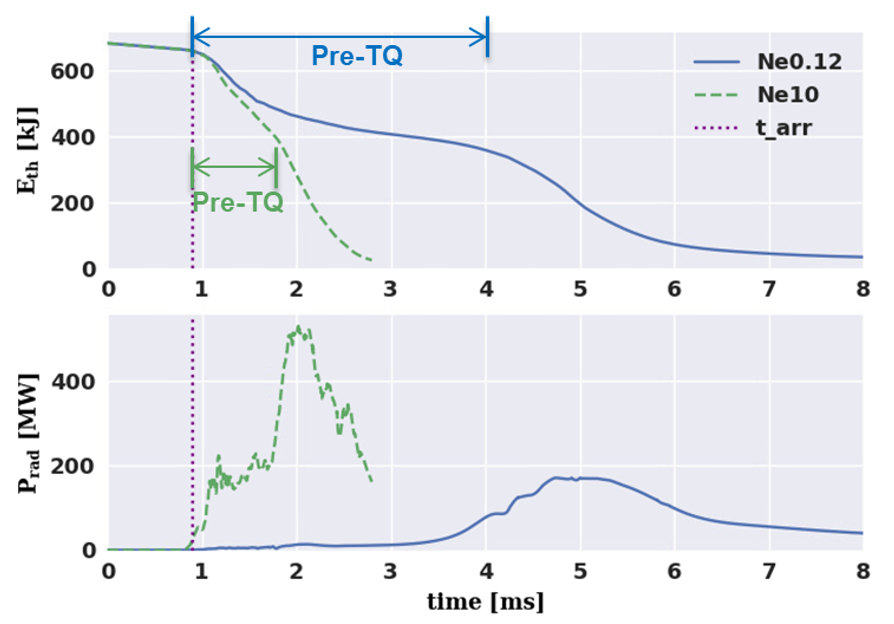}
\caption{Time evolution of total thermal energy and radiation power with different neon fractions 0.12\% and 10\% for cases with $0.1\cdot\chi_{\|,SH}$. The time of fragment arrival $t_{arr}$ is marked by the purple vertical dotted line. The transition from the pre-TQ phase to the TQ phase is identified by the sudden steepening of the thermal energy decay at around 2 and 4~ms, respectively. Accordingly, the pre-TQ duration is defined as the interval between the $t_{arr}$ and this characteristic inflection point.}
\label{fig_ne_th}
\end{figure}

\begin{figure}[htb!]
\centering
\includegraphics[width=.9\textwidth]{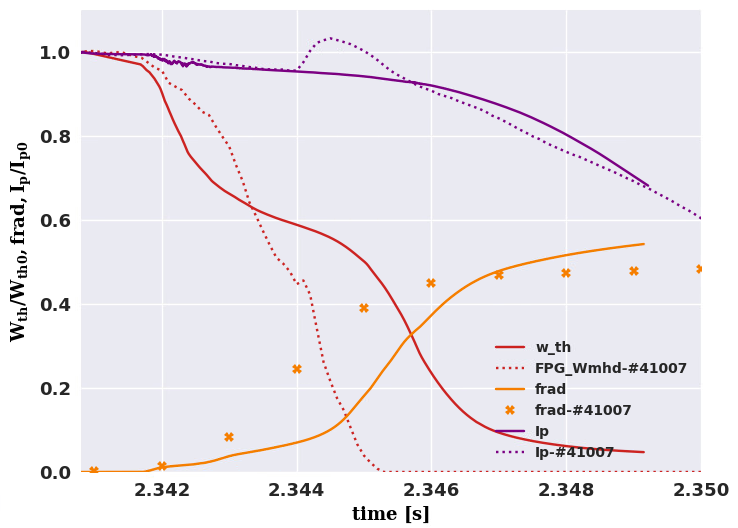}
\caption{Time evolution of thermal energy, plasma current and radiation fraction for injections of 0.12\% neon pellets with $0.1\cdot\chi_{\|,SH}$, in comparison with the corresponding experiment discharge \#41007 (0.17\% neon, pellet speed 248~m/s, $25^\circ$ shatter angle). The thermal energy and plasma current are normalized to the values at t=0 for a clear comparison. The `X' marker shows the radiation fraction from experimental data~\cite{paul2025nf_frad}.}
\label{fig_exp_41007}
\end{figure} 

\begin{figure}[htb!]
\centering
\includegraphics[width=.75\textwidth]{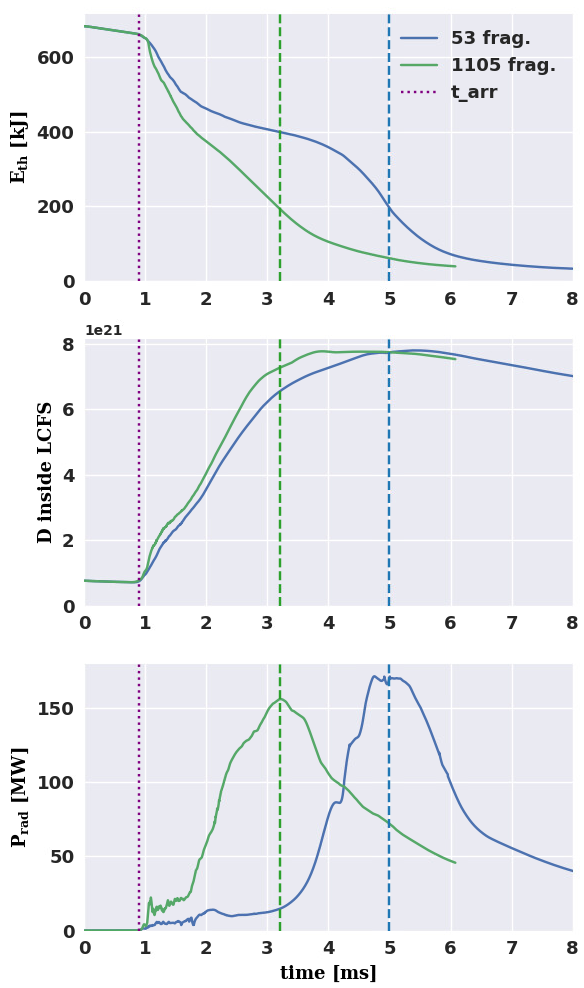}
\caption{Time evolution of total thermal energy, assimilated deuterium atoms and radiation power with different fragment sizes, i.e. 53 fragments and 1105 fragments with 0.12\% neon fraction and $0.1\cdot\chi_{\|,SH}$. The time of fragment arrival $t_{arr}$ is marked by the purple vertical dotted line. The times at the end of TQ are annotated by the vertical dashed lines with different colors, which are in line with the radiation peaks.}
\label{fig_size}
\end{figure}

    The influence of neon content is examined in the first place by introducing a trace neon case with a neon fraction of 0.12\%. To isolate the effect of neon fraction, we assume the fragment configuration is identical, as previously described in figure~\ref{fig_dist}. Figure~\ref{fig_ne_th} presents the time evolution of total thermal energy and radiation power for the two neon fractions, 0.12\% and 10\%. From the thermal energy evolution, it is clear that the two-stage cooling behavior reported in our previous study~\cite{tang2025augspi} persists: the first stage is dominated by convective and conductive transport induced by magnetic field stochastization at the plasma edge, while the second stage is primarily driven by radiation. However, with the reduced $\chi_{\parallel}$, the first-stage cooling becomes much milder compared to the previous case with $\chi_{\parallel,SH}$. After approximately a 20\% reduction in thermal energy during the first stage, the two curves begin to diverge. Notably, a more pronounced pre-TQ phase is observed compared to the $\chi_{\parallel,SH}$ cases, as evidenced by the slope of the thermal energy decay, as well as the radiation peak. This is now in line with neon doped pellets from previous experimental observations~\cite{Heinrich_2025_PhD,Jachmich_2023_EPS,Jachmich_2024_EPS} and INDEX simulations~\cite{ansh_augspi} (see also figure 6.16 in the thesis by P.~Heinrich~\cite{Heinrich_2025_PhD}). The resulting pre-TQ durations, approximately 1~ms for 10\% neon and 3~–~4~ms for 0.12\% neon, fall well within the experimentally observed range~\cite{tang2025augspi,Heinrich_2025_PhD, Heinrich_2026_disr_evo}. Note that the estimation of pre-TQ in experiments, i.e. interval between the main fragments arrival to the current dip, is slightly different from our measurements, because, as we mentioned before, the current spike is not reproduced. Apart from the extended pre-TQ duration, other key metrics remain largely consistent with previous results. At lower neon fractions, the TQ process proceeds more slowly, resulting in a longer ablation duration and consequently higher electron density assimilation. This enhanced assimilation may be beneficial for runaway suppression, as the increased post-TQ density raises collisionality and mitigates runaway electron generation~\cite{Hollmann2015}.

    Similar to figure~\ref{fig_exp_40673}, the thermal energy, plasma current and radiation fraction for the case with $0.1\cdot\chi_{\parallel,SH}$ and 0.12\% neon are compared with corresponding experimental measurements in figure~\ref{fig_exp_41007}. Overall, the results of trace neon case can also match with the experiments relatively well. The two-stage cooling behavior observed in the simulations is also evident in the experimental data. However, the thermal energy loss during the first stage remains slightly overestimated compared to the experiments. This suggests that the initial MHD dynamics may still be overestimated. By the end of the simulation, both the radiation fraction and the plasma current decay agree well with the experimental observations, indicating a reasonable level of neon assimilation in the simulation.  

    Based on the trace neon (0.12\%) case with reduced $\chi_{\parallel}$, we re-assessed the effect of fragment size by adding one extra case with 1105 fragments, whose speed and size distributions are shown in figure~\ref{fig_dist}. The speed distribution is kept comparable to that of the 53 fragment case, while the size distribution differs, allowing us to isolate the influence of fragment size.

    Figure~\ref{fig_size} shows the time evolution of total thermal energy, assimilated deuterium atoms and radiation power for the two trace neon (0.12\%) cases with different fragment sizes. The larger total surface area of the smaller fragments enhances the ablation rate, leading to higher assimilation during the initial phase, consistent with our previous results~\cite{tang2025augspi}. However, the longer pre-TQ duration associated with the larger fragments allows more time for material ablation. Consequently, the total assimilation at the end of the TQ is slightly higher for the larger fragments, accompanied by a relatively stronger radiation peak. Compared to previous results~\cite{tang2025augspi}, the pellet velocity is relatively lower here, so that the fragment traversal time through the plasma exceeds the TQ timescale, thereby eliminating the effect of fragments leaving the plasma domain before the TQ is completed. 

\begin{figure}[htb!]
\centering
\includegraphics[width=1.\textwidth]{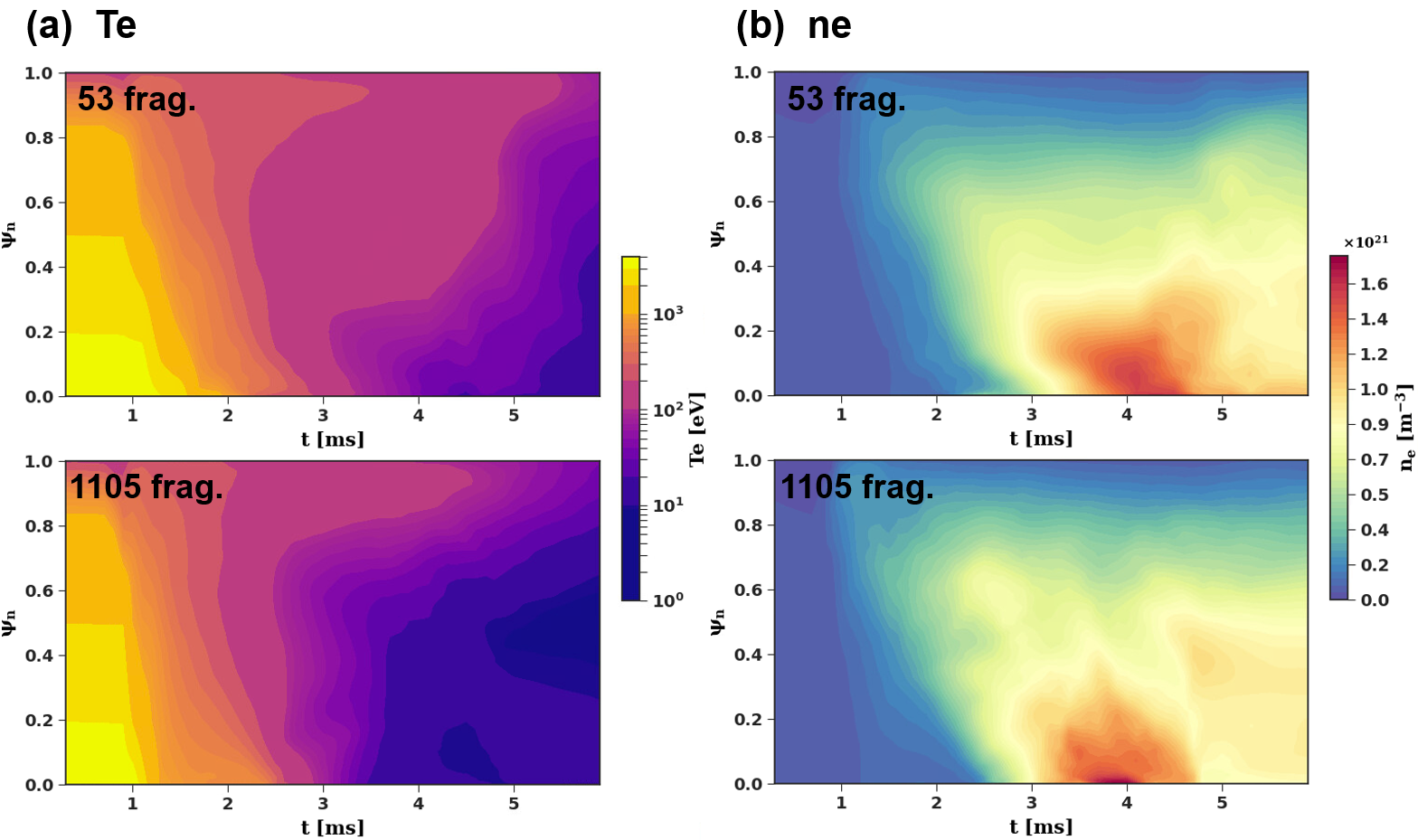}
\caption{Flux-averaged (a) electron temperature $T_e$ profile and (b) electron density $n_e$ evolution for the cases with different fragment sizes, using $0.1\cdot\chi_{\|,SH}$ and 0.12\% neon fraction. }
\label{fig_size_prof}
\end{figure} 

    Flux-averaged electron temperature $T_e$ profile and electron density $n_e$ evolution for the two above cases are illustrated in figure~\ref{fig_size_prof}. Both the cooling front and ablation front of the smaller fragments advance more rapidly than those of the larger fragments. This behavior arises from the velocity distribution, which includes more high-speed fragments for the smaller fragment configuration, as well as from the larger total surface area that accelerates ablation. As a result, the TQ duration is longer for the larger fragments, and a more pronounced density rise in the core region can be observed in the $n_e$ profile evolution. It is worth noting that, in experiments, the pre-TQ duration for smaller fragments with trace neon fraction tends to be even longer than that for larger ones~\cite{Heinrich_2025_PhD}. Despite this longer pre-TQ duration, the neon assimilation inferred from the CQ rate is likely lower for smaller fragments~\cite{Heinrich_2025_PhD,Jachmich_2023_EPS,Heinrich_2026_disr_evo}, suggesting that smaller fragments may encounter penetration limitations. The current model does not incorporate the rocket effect~\cite{rocketprl}, which can be particularly important for small fragments. Their smaller mass makes them more susceptible to this effect, potentially expelling them from deeper penetration into the plasma. Future work will focus on implementing this effect to improve the accuracy of penetration modeling.
        
\section{Summary and discussion}\label{sec4}

    In this study, using the JOREK code, we advanced the 3D non-linear MHD simulations of SPI in AUG H-mode plasmas towards quantitative experimental validation and interpretation, focusing on the impact of reduced parallel thermal diffusivity on disruption dynamics. Applying a flux-limited thermal diffusivity model in a simplified manner by reducing the Spitzer–Härm diffusivity by one order of magnitude, we achieved an improved quantitative agreement with experiments in terms of pre-TQ durations and radiation fractions without compromising the experiment agreement of other aspects of the simulation. 
    
    The reduced thermal diffusivity mitigates the overestimation of parallel heat transport present in previous simulations, leading to a weaker MHD response and a more gradual cooling front that aligns closely with experimental observations. The modification also results in an increased radiation fraction consistent with diagnostic measurements, indicating more realistic thermal energy losses through conductive and convective channels.

    Building upon this improved baseline, we re-examined the influence of neon content and fragment size. Simulations with different neon fractions revealed that the two-stage cooling process—initially dominated by conduction and convection, followed by radiation—persists with the flux-limited $\chi_{\parallel}$. The pre-TQ duration decreases significantly for neon fractions from 0.12\% to 10\%, aligning with the experimentally observed range (1–4 ms). Furthermore, fragment size studies of trace neon fraction (0.12\%) demonstrated that although smaller fragments enhance early ablation through increased surface area, larger fragments yield slightly higher overall assimilation due to longer pre-TQ duration. For the trace neon cases, the assimilation of smaller fragments in the experiment is significantly lower than that of large fragments, exhibiting a stronger influence than that observed in simulations. This difference is very likely due to the rocket effect.

    Overall, this work emphasizes the critical role of flux-limited parallel heat transport in realistic SPI disruption simulations. Incorporating reduced $\chi_{\parallel}$ improves agreement with experimental dynamics and provides a more reliable foundation for optimizing SPI configurations for ITER’s disruption mitigation system. The present JOREK simulation can reproduce the experiments to a significant extent and will be utilized in future experimental interpretation. Future efforts will focus on implementing more sophisticated models, including more realistic treatment of the flux-limited parallel conduction model, and the pellet rocket effect to capture small fragment penetration more accurately. 
    
\section*{Acknowledgements}
    We sincerely thank Mengdi Kong and Daniele Bonfiglio for useful discussions. We are very grateful for the CPU time generously provided by the HAWK supercomputer operated by the High Performance Computing Center (HLRS) in Stuttgart, Germany. Part of this work has been carried out within the framework of the EUROfusion Consortium, funded by the European Union via the Euratom Research and Training Programme (Grant Agreement No 101052200 — EUROfusion). Views and opinions expressed are however those of the author(s) only and do not necessarily reflect those of the European Union or the European Commission. Neither the European Union nor the European Commission can be held responsible for them. Part of this work has been carried out in collaboration with the ITER Organization. The views and opinions expressed herein do not necessarily reflect those of the ITER Organization. This work receives funding from the ITER Organization under contract IO/20/IA/43-2200. The ASDEX-Upgrade SPI project has been implemented as part of the ITER DMS Task Force programme. The SPI system and related diagnostics have received funding from the ITER Organization under contracts IO/20/CT/43-2084, IO/20/CT/43-2115, IO/20/CT/43-2116.
\section*{References}
\bibliography{paper}
\end{document}